\documentclass{gmee2012}

\input epsf
\usepackage{graphicx}
\usepackage{amsmath}   % From the American Mathematical Society
\usepackage[retain-explicit-plus]{siunitx}
\usepackage{color}
\usepackage{subfigure}
\usepackage{booktabs}
\usepackage{cite}

\newcommand{\revision}[1]{#1}

\begin{document}

%+++++++++++++++++++++++++++++++++++++++++++
\title{\LARGE On the calibration \\ of direct-current current transformers}
%\author{\authorblockN{J. Clerk Maxwell}
%\authorblockA{School of Electrical and\\Computer Engineering\\
%Somewhere Institute of Technology\\
%City, State 54321--0000\\
%Email: maxwell@curl.edu}
%\and
%\authorblockN{Michael Faraday}
%\authorblockA{(List authors on this line using 12 point Times font\\ - use a second line if necessary)\\
%Microwave Research\\
%City, State/Region, Mail/Zip Code, Country\\
%\and
%\authorblockN{Andr\'e M. Amp\`ere \\ }
%\authorblockA{Starfleet Academy\\
%San Francisco, CA 96678-2391\\
% avoiding spaces at the end of the author lines is not a problem with
% conference papers because we don't use \thanks or \IEEEmembership

\author{Luca Callegaro,
        Cristina Cassiago,
        and~Enrico Gasparotto
\thanks{Luca Callegaro, Cristina Cassiago and Enrico Gasparotto are with the Electromagnetism Division of the Istituto Nazionale di Ricerca Metrologica (INRIM), Strada delle Cacce 91, 10135 Torino, Italy, e-mail: \texttt{l.callegaro@inrim.it}}% <-this % stops a space
\thanks{In press on IEEE Trans. Instr. Meas.}}

%\specialpapernotice{(Invited Paper)}


\maketitle

%\IEEEpeerreviewmaketitle

\begin{abstract}
Modern commercial direct-current current transformers (DCCT) can measure currents up to the \si{\kilo\ampere} range with accuracies better than \num{1E-5}. We discuss here a DCCT calibration method and its implementation with commercial instruments typically employed in low resistance calibration laboratories. The primary current ranges up to \SI{2}{\kilo\ampere}; in the current range below \SI{100}{\ampere} the calibration uncertainty is better than \num{3E-7}. An example of calibration of a high-performance DCCT specified for primary currents measurement up to \SI{900}{\ampere} is discussed in detail.
\end{abstract}

%\IEEEoverridecommandlockouts
%\begin{keywords}
%Current measurement; metrology; calibration; transducers; magnetic sensors.
%\end{keywords}

% for peerreview papers, inserts a page break and creates the second title.
% Will be ignored for other modes.
%\IEEEpeerreviewmaketitle

\section{Introduction}
Direct-current current transformers (DCCT) are the most accurate dc high-current sensors commercially available \cite{Ripka2010}, reaching specified relative accuracies in the \num{E-5} range and integral nonlinearities below \num{E-6}. The verification of such high performances and the calibration of the DCCT ratio require metrological facilities capable of handling high currents, with high accuracy and automated operability~\cite{Zhu1998,Fernqvist2002,Fernqvist2003,Hudson2008}.

Ultimate current ratio accuracy is achieved in cryogenic current comparators (CCC) \cite{Williams2011}. In a CCC, ratio accuracy is obtained by constraining the magnetic flux (generated by the current being compared) within superconducting shields. An extremely high sensitivity is achieved with a superconducting quantum interference device (SQUID) flux sensor. Even though CCCs capable of handling currents up to \SI{100}{A} have been realized \cite{Williams1999}, these devices are research instruments not available in calibration laboratories.

Ferromagnetic-core, room-temperature current comparators (CC) are current ratio devices which can achieve ratio errors lower than \num{E-7} \cite{Moore1988}, and can be self-calibrated through step-up procedures \cite{Shao2010,Zhao2010} with similar levels of uncertainty. Thus, a CC can be employed as current ratio standard in a DCCT calibration setup. Although complex and expensive instruments, high-current CC are common in electrical calibration laboratories, since they are part of commercial resistance ratio bridges employed for the measurement of low-value resistors. These instruments include also current sources, detectors, and firmware for automated operation. 

\revision{The calibration of the DCCT ratio with a reference current ratio standard (possibly having a different nominal ratio) can be performed by different methods. Recent papers \cite{Jeckelmann2004, Rietveld2013} describe a method based on the comparison of the voltages developed by the secondary currents of the devices being compared on calibrated resistance standards.}

Here we present a simple method that allows the calibration of the ratio of a DCCT by using commercial components, originally designed for the calibration of low-value resistors. \revision{This method does not require calibrated resistance standards; }the accuracy, dependent on the primary current, is better than \num{3E-7} for currents below \SI{100}{A}. An example of calibration of a DCCT having a $1500:1$ nominal ratio for currents up to \SI{900}{\ampere} is reported.

The implementation is being employed in the EURAMET.EM-S35 High DC current ratio supplementary comparison~\cite{EURAMET1217}, in which INRIM acts as co-pilot laboratory. 
\section{Calibration method}
\label{sec:method}
\begin{figure}[tb]
	\centering
	\includegraphics[width=0.6\linewidth]{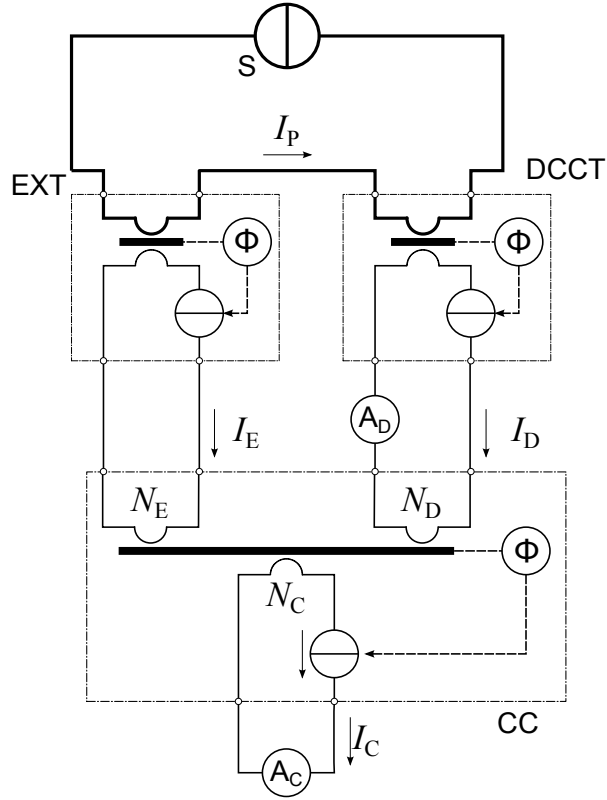}
 	\caption{Schematic diagram of the DCCT calibration setup. S is a high-current source, EXT is an automated current range extender and CC is a dc current comparator. The ammeter $\textup{A}_\textup{D}$ monitors the DCCT output \revision{$I_\textup{D}$}, while $\textup{A}_\textup{C}$ measures the current $I_\textup{C}$ which is the input quantity of the measurement model \eqref{eq:balance}.   \label{fig:CalScheme}}%
\end{figure}
\begin{figure}[tb]
	\centering
	\includegraphics[width=0.5\linewidth]{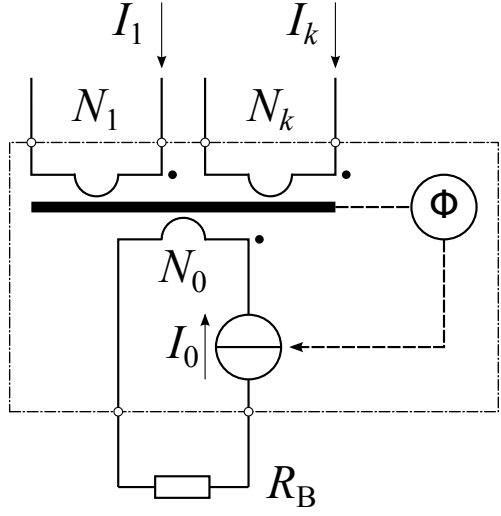}
 	\caption{\revision{The operating principle of the instruments employed in the calibration setup. $m+1$ windings, with $N_k$ $(k=0\ldots m)$ turns each, are wound around a ferromagnetic core (black thick line). The flux detector $\Phi$, through an automatic control (not shown), steers the current source $I_0$ until $\Phi=0$; a generic load $R_\textup{L}$ can be connected to the source.}   \label{fig:Principle}}%
\end{figure}

\revision{Fig. \ref{fig:CalScheme} shows the schematic diagram of the calibration setup which includes three current ratio devices: the DCCT under calibration, an automated current range extender EXT and a current comparator ratio bridge CC.}
\revision{\subsection{Operation of current ratio devices}}
The operation of the three current ratio devices, \revision{sketched in Fig.~\ref{fig:Principle}}, is based on the same principle.

\revision{$m+1$ windings are wound around a ferromagnetic core. Each winding $k$ has $N_k$ $(k=0\ldots m)$ turns, and a current $I_k$ flows through it. The magnetic flux $\Phi$ in the core is given by $\mathcal{R} \Phi = \sum\revision{_{k=0}^m} N_k I_k $, where $\mathcal{R}$ is the core magnetic reluctance. $\Phi$ is measured by a fluxgate detector~\cite{Macmartin1966,Odier2004,Ripka2010} whose output constitutes the error signal of a feedback control. The output of the control drives current source $I_0$, connected to winding $0$, to null the flux. The condition $\Phi = 0$ yields the ampere-turns balance equation $\sum_{k=0}^m N_k I_k = 0$. }

\revision{In normal operating conditions,} DCCT and EXT have only two ($m=1$) active windings. The output $I_0$ of the controlled current source constitutes the device output current; hence, the current $I_1$ is scaled down with the turns ratio $n_{1,0} = N_1/N_0$ as $I_0 = n_{1,0} I_1$. 

In the CC, instead, more windings ($m \geq 2$) are simultaneously active; the currents \revision{$I_k$ ($k=0\ldots m$)} are compared, weighted by their respective turn numbers $N_k$; the measurement of $I_0$ \revision{gives the CC reading, that is, the residual unbalance between the currents $I_1 \ldots I_m$ to be compared.} 
\revision{\subsection{Circuit description}}
The input windings of both DCCT and EXT are connected in series and driven by the primary current $I_\textup{P}$ generated by the high-current dc source S. The DCCT and EXT output currents are respectively $I_\textup{D}=G_\textup{D} I_\textup{P}$ and $I_\textup{E} = G_\textup{E} I_\textup{P}$, where $G_\textup{D}$ is the DCCT current gain (that is, the measurand) and $G_\textup{E}$ is the EXT current gain. 

$I_\textup{D}$ and $I_\textup{E}$ are connected to two input windings of CC, each having $N_\textup{D}$ and $N_\textup{E}$ turns. $I_D$ is also measured by a high-accuracy ammeter $\textup{A}_\textup{D}$.

The CC compensation current $I_\textup{C}$, linked to the CC winding with $N_\textup{C}$ turns, is measured by the ammeter $\textup{A}_\textup{C}$; when operating properly, the CC balance equation is 
\begin{equation}
	\label{eq:balCC}
	N_\textup{E} I_\textup{E} + N_\textup{D} I_\textup{D} + N_\textup{C} I_\textup{C} = 0.
\end{equation}
In \eqref{eq:balCC}, the sign of turn numbers $N_x$ can be either positive or negative and is set by the winding direction. 

When in all current ratio devices each core flux is drawn to zero by the corresponding automated control, the balance equation of the whole circuit becomes
\begin{equation}
	\label{eq:balance}
	G_\textup{D} = \frac{I_\textup{C}}{I_\textup{P}} \frac{N_\textup{C}}{N_\textup{D}} - G_\textup{E} \frac{N_\textup{E}}{N_\textup{D}}. 
\end{equation} 

\revision{\subsection{Measurement model}}
\revision{To derive an accurate measurement model, two major nonidealities of the devices employed should be considered:
\subsubsection{Offset}  All instruments based on the fluxgate technique suffer from a certain degree of dc offset, caused by the magnetization hysteresis and relaxation of the ferromagnetic core. This offset, of the order of \SI{10}{\micro\ampere} per unit input turn \cite{Macmartin1966}, depends on temperature, measurement history and time drifts. To compensate this offset, the reading $I_\textup{C}$ in \eqref{eq:balance} is substituted with $\Delta I_\textup{C} = I_\textup{C} - I_\textup{C0}$, where $I_\textup{C}$ is the reading taken at the nominal primary current $I_\textup{P}^\textup{(n)}$ of interest, and $I_\textup{C0}$ is the reading with null primary current, $I_\textup{P} = 0$.
\subsubsection{Ratio errors}
The actual current ratios of CC can differ from the corresponding turn ratios. We call $n_\textup{CD}$ and $n_\textup{ED}$ the current ratios of which $N_\textup{C}/N_\textup{D}$ and $N_\textup{E}/N_\textup{D}$ are the corresponding nominal turn ratios.

Taking into account the above nonidealities, \eqref{eq:balance} can be rewritten as
\begin{equation}
	\label{eq:measurement_model}
	G_\textup{D} = \frac{\Delta I_\textup{C}}{I_\textup{P}} \, n_\textup{CD} - G_\textup{E} \, n_\textup{ED}, 
\end{equation} 
}

The relative gain error $\delta G_\textup{D}$ with respect to the nominal gain $G_\textup{D}^\textup{(n)}$ is
\begin{equation}
	\label{eq:deltaGD}
	\delta G_\textup{D} = \frac{G_\textup{D} - G_\textup{D}^\textup{(n)}}{G_\textup{D}^\textup{(n)}}.
\end{equation}

\section{Implementation}
\label{sec:implementation}
\begin{figure}[tb]
	\centering
	\includegraphics[width=0.6\linewidth]{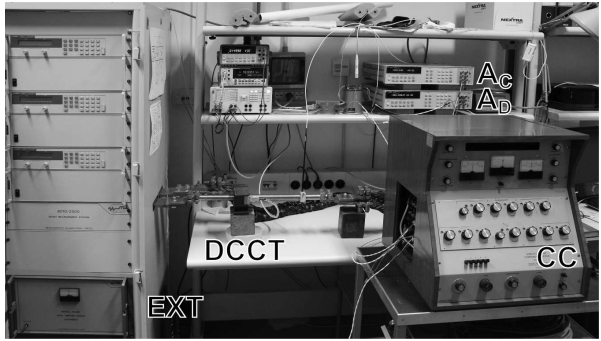}%
 	\caption{\revision{A photo of the implementation of the calibration method. See Sec.~\ref{sec:implementation} and Fig.~\ref{fig:CalScheme} for the identification of the labels.  \label{fig:photosetup}}}%
\end{figure}
Fig.~\ref{fig:photosetup} shows an implementation of the schematic diagram of Fig.~\ref{fig:CalScheme}. It employs the following instrumentation:
\begin{description}
\item[DCCT] \hspace{3mm} The device under test, for which the results reported in Sec.~\ref{sec:results} were obtained, is a LEM mod.~ITN~900-S~ULTRASTAB high-performance current transducer~\cite{Teppan2010}. It handles primary currents $|I_\textup{P}|\leq  \SI{900}{\ampere}$ with a nominal current ratio $G_\textup{D} = 1/1500$. The specified accuracy is better than \num{2E-5} (including offset), the linearity better than \num{1E-6}, maximum load resistance~\SI{2.5}{\ohm}. Fig.~\ref{fig:photo} shows the DCCT mounted on the primary busbar.
\begin{figure}[ht]
	\centering
	\includegraphics[width=\linewidth]{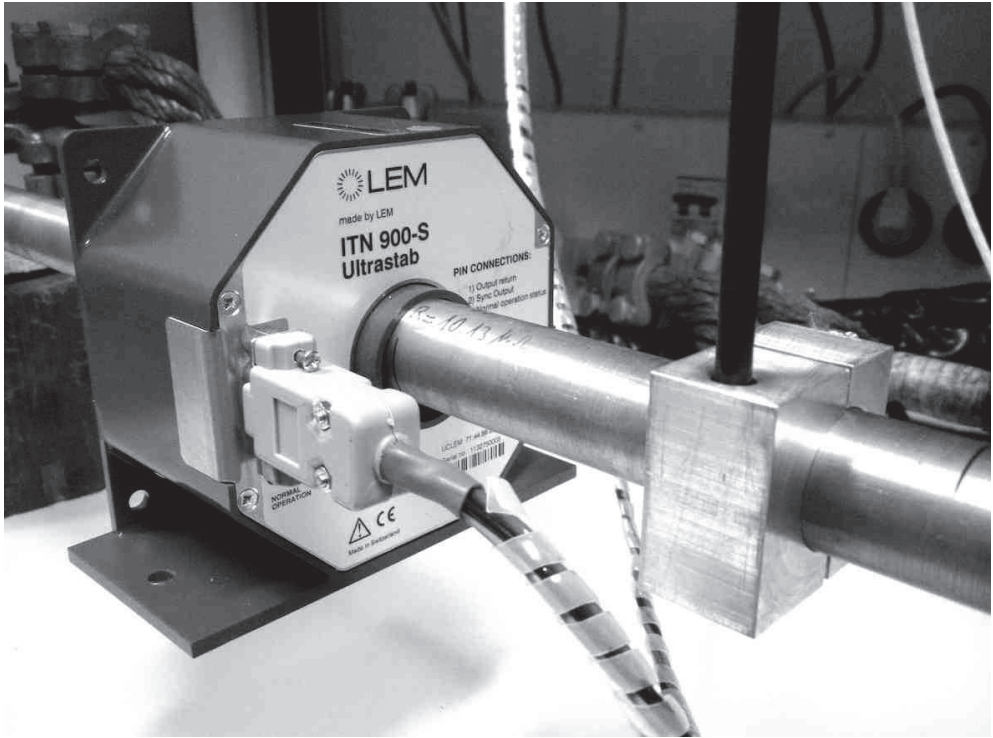}%
 	\caption{The DCCT under calibration mounted on the primary current busbar. The aluminum block in foreground embeds a Pt100 sensor to monitor the temperature of the primary current busbar.  \label{fig:photo}}%
\end{figure}
\item[CC] Guildline mod.~9920 direct current comparator~\cite{Macmartin1966}. This instrument is particularly versatile since it provides several fixed windings having decadic ($1$ to $1000$) number of turns and one winding with an adjustable number of turns through decade rotary switches; moreover, it allows a full reconfiguration of the connections between the windings and the internal electronics. The settings used in the calibration of the particular DCCT under test are: $N_\textup{D} = -100$ (fixed winding), $N_\textup{E}=150$ (decade winding), and $N_\textup{C}=1$ in order to achieve the highest sensitivity in the measurement of $I_\textup{C}$.
\item[EXT] Two different extenders were employed, depending on $I_\textup{P}$:
\begin{itemize}
\item Measurement International mod. MI~6011B range extender. Primary current $|I_\textup{P}| \leq  \SI{100}{\ampere}$, nominal ratio $1/1000$, relative accuracy $<\num{1E-7}$.
\item Measurement International mod. MI~6012M range extender. $|I_\textup{P}| \leq  \SI{2}{\kilo\ampere}$, nominal ratio $1/1000$, relative accuracy $<\num{2E-6}$.
\end{itemize}
The above specifications were validated in the standard operating setup for low-valued resistor measurements~\cite{Kraft2012}.  
\item[S] Two different sources were employed, depending on the primary current $I_\textup{P}$:
\begin{itemize}
\item Measurement International MI~6100A linear dc power supply, for $|I_\textup{P}| < \SI{100}{\ampere}$. Current reversal is achieved with a  switch internal to MI~6011B. 
\item Agilent mod.~6680 (two items in parallel) for $|I_\textup{P}| < \SI{1750}{\ampere}$. Current reversal is achieved with a Measurement International mod.~6025 pneumatic switch.
\end{itemize}
\item[$\textup{A}_\textup{D}$] Agilent mod. 3458A multimeter in dc voltage mode, measuring the voltage drop on a Tinsley mod.~1659 $\SI{1}{\ohm}$ standard resistor.
\item[$\textup{A}_\textup{C}$] Agilent mod. 3458A multimeter in dc current mode, \SI{100}{\milli\ampere} range. 
\end{description}
The DCCT and busbar temperatures are monitored with two Pt100 platinum temperature sensors (see Fig.~\ref{fig:photo}) read by a Fluke mod.~1529 CHUB E-4 thermometer. 
\section{Results}
\label{sec:results}
\begin{figure}[t]
	\centering
	\includegraphics[width=0.6\linewidth]{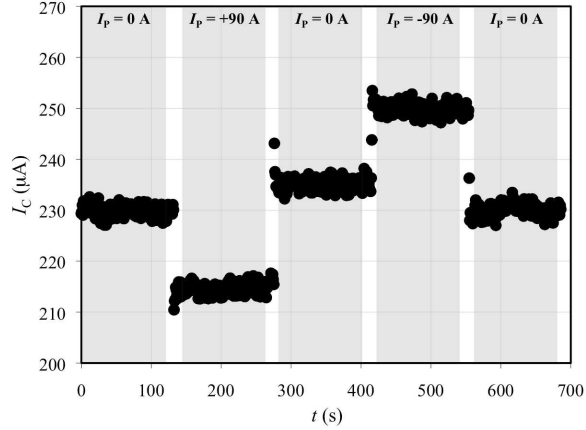}%
 	\caption{Time recording of the compensation current $I_\textup{C}$ for different primary currents $I_\textup{P}$.   \label{fig:Ic}}%
\end{figure}

\revision{After a warming-up period of about \SI{1}{\hour} at $I_\textup{P}=+I_\textup{P}^\textup{(n)}$, $I_\textup{P}$ is repeatedly cycled between values $0$, $+I_\textup{P}^\textup{(n)}$, $0$, $-I_\textup{P}^\textup{(n)}$ (ending the whole cycle sequence with $I_\textup{P}=0$)}.\footnote{\revision{It has been found that, for the particular DCCT being tested and for $I_\textup{P}$ near fullscale, the current must be ramped up smoothly to allow the DCCT automatic shutdown.}}  The reading $I_\textup{C}$ is continuously recorded. Fig.~\ref{fig:Ic} shows a time series of $I_\textup{C}$ readings corresponding to an $I_\textup{P}$ cycle.  
For each value of $I_\textup{P}$, after transients have died out, a time average $\overline{I_\textup{C}}(I_\textup{P})$ is computed (see gray bands in Fig.~\ref{fig:Ic}). 

The quantity $\Delta I_\textup{C}$ to be employed in Eq.~\eqref{eq:balance} is computed as  $\Delta I_\textup{C}(I_\textup{P}^\textup{(n)}) = \overline{I_\textup{C}}(I_\textup{P}^\textup{(n)}) - (\overline{I_\textup{C}'}(0) + \overline{I_\textup{C}''}(0))/2$, where $\overline{I_\textup{C}'}(0)$ and $\overline{I_\textup{C}''}(0)$ are the zero readings respectively preceding and succeeding $\overline{I_\textup{C}}(I_\textup{P}^\textup{(n)})$ in the time series. 

\begin{figure}[t]
	\centering
	\includegraphics[width=0.6\linewidth]{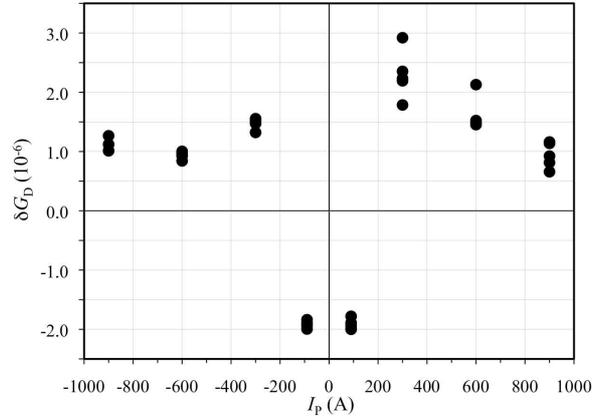}%
 	\caption{The relative gain error $\delta G_\textup{D}$ of the DCCT under calibration, for different nominal primary current values $I_\textup{P}^\textup{(n)}$. Each dot corresponds to half of the measurement cycle shown in Fig.~\ref{fig:Ic}. For each $I_\textup{P}^\textup{(n)}$, five measurement results are reported. The average value and its uncertainty are given in Tab.~\ref{tab:results}.  \label{fig:DeltaG}}%
\end{figure}

From each $\Delta I_\textup{C}(I_\textup{P}^\textup{(n)})$, the absolute $\Delta G_\textup{C}(I_\textup{P}^\textup{(n)})$ and relative $\delta G_\textup{C}(I_\textup{P}^\textup{(n)})$ errors are computed. Fig.~\ref{fig:DeltaG} graphically shows the values $\delta G_\textup{D}$ corresponding to each $I_\textup{P}^\textup{(n)}$ measurement cycle.

Tab.~\ref{tab:results} reports the estimates for $G_\textup{D}$ and $\delta G_\textup{D}$ of the DCCT under measurement, together with the corresponding \SI{95}{\percent} expanded uncertainties, for several primary current $I_\textup{P}^\textup{(n)}$ values.
\begin{table}[t]
	\centering
	\caption{DCCT gain $G_\textup{D}$ for different primary currents $I_\textup{P}$. The relative deviation $\delta G_\textup{D}$  from nominal ratio is also reported. \label{tab:results}}
	\begin{tabular}{ccclll}
		\toprule
		$I_\textup{P}^\textup{(n)}$ & Supply & EXT & $G_\textup{D}$ & $\delta G_\textup{D}$ & $U(\delta G_\textup{D})$ \\
		$\si{\ampere}$ &  & & $\times\num{E-4}$ & $\times\num{E-6}$ & $\times\num{E-6}$ \\
		\midrule		
		\num{+90}		&6100A		&6011B	&\num{6.6666539(16)}&\num{-1.92}	&\num{0.24}\\
		\num{-90}		&6100A		&6011B	&\num{6.6666539(16)}&\num{-1.92}	&\num{0.23}\\
		\num{+300}		&6680A		&6012M	&\num{6.666682(16)}&\num{+2.3}	&\num{2.3}\\
		\num{-300}		&6680A		&6012M	&\num{6.666676(15)}&\num{+1.5}	&\num{2.3}\\
		\num{+600}		&6680A		&6012M	&\num{6.666677(15)}&\num{+1.6}	&\num{2.3}\\
		\num{-600}		&6680A		&6012M	&\num{6.666676(15)}&\num{+0.9}	&\num{2.3}\\
		\num{+900}		&6680A		&6012M	&\num{6.666672(15)}&\num{+0.9}	&\num{2.3}\\
		\num{-900}		&6680A		&6012M	&\num{6.666673(15)}&\num{+1.1}	&\num{2.3}\\
		\bottomrule
\end{tabular}
\end{table}

As an example, the uncertainty budget for the calibration of $G_\textup{D}$ at $I_\textup{P}=\SI{+90}{\ampere}$ is given in Tab.~\ref{tab:unc90A}, where it can be appreciated that the main contributions to the measurement uncertainty are due to the instability of $I_\textup{C}$ and the EXT and CC current ratios \revision{$G_\textup{E}$ and $n_\textup{ED}$}.
\begin{table*}[t]
	\small
	\centering
	\caption{Uncertainty budget for $G_\textup{D}$, at $I_\textup{P}^\textup{(n)} =  \SI{+90}{\ampere}$.   \label{tab:unc90A}}
	%\resizebox{\linewidth}{!}{
	\tabcolsep=0.1cm
	\begin{tabular}{lccccl}
		\addlinespace[5pt]
		\toprule
		Quantity & $X$ & $u(X)$ & contrib. to $u(G_\textup{D})$ & type & \revision{note}\\
		\midrule
		$\Delta I_\textup{C}$ 			& \SI{-17.2}{\micro\ampere} 	 & \SI{0.36}{\micro\ampere} 	& \num{2.7E-11}		& A	& \revision{Standard deviation of the mean, taken over \num{5} $I_\textup{P}$ cycles} \\
		$I_\textup{P}$ 	 				& \SI{90}{\ampere}			 & \SI{90}{\milli\ampere}	 & \num{1.3E-12}   	& B & \revision{$\textup{A}_\textup{D}$ readings (bound on maximum error)}\\
		\revision{$n_\textup{CD}$}		& \num{-6.6666667E-03}		 & \num{6.7E-10} 	 		 & $<\num{1E-13}$	& B & \revision{CC manufacturer's specifications}\\
		\revision{$n_\textup{ED}$} 		& \num{-6.6666667E-01}		 & \num{6.7E-08} 			 & \num{6.7E-11}  	& B & \revision{CC manufacturer's specifications}\\ 
		$G_\textup{E}$ 					& \num{+1.0000000E-03}		 & \num{5.8E-11} 			 & \num{3.8E-11}   	& B & \revision{EXT (MI 6011B) manufacturer's specifications}\\
		\midrule
		$G_\textup{D}$					& \num{+6.6666539E-04}		 & \num{8.1E-11}		     	& 	 				 &  & \\
		$\delta G_\textup{D}$			& \num{-1.92E-6}			 	& \num{1.2E-7}  				&					 &  &\\
		\revision{$U(\delta G_\textup{D})$} 		& 					& \num{2.4E-7} 		   		&   					 &  &\revision{Expanded uncertainty, \SI{95}{\percent} coverage probability}\\		
		\bottomrule
	\end{tabular}
	%} %resizebox
\end{table*}
\section{Conclusions}
The proposed setup allows the calibration of the ratio of a DCCT with accuracies in the \num{E-6} range or better. The proposed implementation, suitable for primary currents up to \SI{2}{\kilo\ampere}, is based on commercial instruments typically employed for the calibration of low-valued resistors, and therefore often available in calibration laboratories. \revision{The implementation is being employed for the participation to the EURAMET.EM-S35 comparison, which is co-piloted by INRIM and the Federal Institute of Metrology (METAS), Switzerland. The travelling standard of comparison is based on a LEM mod.~IT~600-S ULTRASTAB transducer; the participants measure $G_\textup{D}$ at primary currents $I_\textup{P} = $\SIlist{\pm 90; \pm 300; \pm 600}{\ampere}. The preliminary characterizations of the travelling standard performed by INRIM and METAS give results which are in agreement within a compound relative uncertainty better than \num{5E-6}. At the present time the results of the comparison are confidential; a full validation of the INRIM method will become available after the publication of the comparison report, expected by the end of 2015.}
\section*{Acknowledgments}
The authors are indebted with Alessandro Mortara, Federal Institute of Metrology (METAS), Switzerland, for fruitful discussions; and with their colleagues Fulvio Francone for help in the construction of the calibration facility, and Massimo Ortolano for having reviewed the manuscript.
\bibliographystyle{IEEEtran}
\bibliography{DCC}

\end{document}